\documentclass[12pt]{article}

\advance\voffset by 4\baselineskip \advance\vsize
by-2
\baselineskip \tolerance=2000
\date{}

\long\def\comment#1\endcomment{}
\def\eop{\hfill$\fbox{}$\medskip}

\def\:{\colon}
\long\def\comment#1\endcomment{}

\usepackage{etoolbox}
\makeatletter
\patchcmd\maketitle{\def\@makefnmark{\rlap{\@textsuperscript{\normalfont\@thefnmark}}}}{}{}{}
\makeatother

\makeatletter
\def\thanksAAffil#1{
  \footnotemarkAAffil\protected@xdef\@thanks{\@thanks%
        \protect\footnotetextAAffil[\the \c@footnoteAAffil]{#1}}%
}
\def\thanksANote#1{%
  \footnotemarkANote%
  \protected@xdef\@thanks{\@thanks%
        \protect\footnotetextANote[\the \c@footnoteANote]{#1}}%
}
\makeatother

\newtheorem{theorem}{Theorem}[section]

\newtheorem{lemma}[theorem]{Lemma}

\begin{document}

\title{An Overview of Some Single Machine Scheduling Problems: Polynomial Algorithms, Complexity and Approximability}

\author{Nodari Vakhania$^{1,}$\thanks{Correspondence: nodari@uaem.mx}, Frank Werner$^2$, Víctor Pacheco-Valencia$^1$\\[0.5ex]  and 
Kevin Johedan Ramírez-Fuentes$^1$\\[1ex] 
$^1$ Centro de Investigación en Ciencias\\ Universidad Autónoma del Estado de Morelos
\\ Cuernavaca 62209, Mexico  \\[1ex]
$^2$ Faculty of Mathematics, Otto-von-Guericke University, \\  39106 Magdeburg, Germany\\[1ex]
}

\maketitle

\noindent
{\bf Abstract:}\\
 Since the publication of the first scheduling paper in 1954, a huge number of works dealing with different types of single machine problems appeared. They addressed many heuristics and enumerative procedures, complexity results or structural properties of certain problems. Regarding surveys, often particular subjects like special objective functions are discussed, or more general scheduling problems were surveyed, where a substantial part is devoted to single machine problems. In this paper we present some results on polynomial algorithms, complexity and approximation issues, where the main focus is on results, which have been published during the last decades in papers, where at least one of the first two authors of this paper was involved. We hope that the reviewed results will stimulate further investigation in related research fields.  

\vspace{4mm}

\vspace{4mm}

\section{Introduction}


{\em Scheduling problems} arise in many real-life applications including 
our everyday life. We commonly desire to optimize our time arranging
different activities, where the time is limited and also the resources
that we use are scarce. We wish to optimize some of our personal criteria
but we also depend on the availability of the latter resources. 
As a simple but very common daily example, consider an apartment
with a single bathroom, where a small family lives. The family 
members gets up at some fixed time, and all of them need to 
spend some time in the bathroom, which is a scarce {\it resource} 
(or machine). Typically such a (renewable) resource may serve 
a single object or {\em job}, a family member in our example. 
Every member of the family needs to leave the apartment at
some fixed time to reach his/her job or school without delay.
This time can be different for different members of the family. 
Here arises a small optimization problem that asks to 
schedule the time interval for each member of the family so as
to meet the deadline of each member of the family. 

Generally, in a scheduling problem we are given {\it jobs} or 
{\it tasks} which are some requests that need to be
performed by some resource, a {\it machine} or a {\it processor}. 
In our small example, bathroom requirements for each member of
the family are jobs and the bathroom is a machine. We can give
a vast amount of examples of jobs and machines in our practical
life; e.g., a job in a factory or a software application
in a computer or on a smart phone, or a lesson at school. 
Here machines, computers, smart phones, classrooms and teachers
are resource examples of. Every jobs requires some  time 
units on a resource, whereas a resource is able to process
at most one job at a time. Indeed, one classroom cannot be used
for two different classes simultaneously or a a teacher cannot
give two or more classes at the same time. There is a restriction
on the whole time window during which a factory or a teacher 
can work. It is a scheduling task to arrange the order for
processing the jobs. Often there is some minimization or
maximization criterion called an {\it objective function}. For
example, a very common criterion is to minimize the maximum
job or machine completion time or the {\it makespan} which is
commonly denoted by $C_{\max}$.

We shall denote the given set of $n$ jobs by $\{1,2,\dots,n\}$. 
The processing time of job
$j$ is denoted by $p_j$. Job $j$ may have some additional
characteristics. For example, the time moment from when 
job $j$ is available to be processed is called its {\it release 
or readiness time} and is commonly denoted by $r_j$. In many 
applications, a job, once completed by the machine, needs to be
delivered to the customer by an independent transportation unit, 
which may not be a scarce resource in a given model. Such a 
 {\it delivery time} (or {\it tail})  of job $j$ is commonly
 denoted by $q_j$. The job delivery takes no machine time (a job
is delivered by an independent agent to a customer), though 
$q_j$ contributes to the completion time of job $j$. Hence, 
the makespan crucially depends on the job delivery times. The {\it due
date} $d_j$ is the desirable completion time of job $j$. Both
above parameters specify the urgency of job $j$: the more
is $q_j$, the more urgent is job $j$; likewise, the less is $d_j$, the
more urgent is job $j$. 

All above parameters are normally non-negative integral numbers. In some 
cases, job $j$ is not allowed to be completed after its due date. Then the
term {\it deadline} instead of ``due date''  is commonly used.
Thus, if deadlines are given, our objective is to find a schedule in
which all jobs are completed before or at their deadlines. With due dates,
our objective will be to minimize or maximize some due date
oriented objective function. Let the {\it lateness} of a job in a
schedule be the difference between its completion time and its due
date. One of the common objective functions is the maximal job
lateness $L_{\max}$ (which we have to minimize). Another common
objective function is the number of jobs completed at or before
their due dates $\sum U_j$ which we have to maximize ($U_j$ is a
0-1 variable indicating whether job $j$ is completed on time or not in the
schedule). We may allow to omit some jobs in the case when 
the total workload is more than the machine capacity. In many 
applications, job due dates are important since, if a job is
not completed by its due date then its benefit or functionality
becomes neglectable. Hence, we may wish to maximize the 
throughput, i.e.,the  number of jobs completed on time by their 
due dates. We may also consider more general criteria
to maximize some profits. The profit of
job $j$ can be expressed by its {\it weight} which is
commonly  denoted  by $w_j$.
Job {\it preemptions} might be allowed or not. In a preemptive
scheduling problem, a job can be interrupted on the machine and
later resumed on the machine. In a
non-preemptive scheduling problem, a job $J_j$, once scheduled on
the machine, should be processed without any interruption until its
full completion on the machine. The machine is called {\it idle} at
some moment $t$ if no job is processed by the machine at that
moment, otherwise it is said to be {\it busy}.

Problem restrictions on the way in which the jobs are to be 
performed by machines define the set of {\it feasible solutions}.
For example, a machine can process at most one job simultaneously; 
the  {\it precedence relations} defined on  the set of jobs, 
restrict the relative order in which the jobs can be scheduled. 
Scheduling problems are sometimes referred to as {\em sequencing 
problems} since the jobs need to be scheduled in a sequential manner
(as no job overlapping is permitted). The precedence relations
between the jobs can be represented by directed graphs which
are often referred to as  {\it precedence (task) graphs}. In 
such graphs, nodes represent jobs, and there
is an arc $(i,j)$ if and only if job $i$ immediately precedes job $j$.

A {\it schedule} prescribes to each job the
starting time on the machine creating in this way a total order 
on the set of jobs by sequencing the jobs on the machine.
Denote by $s^S_j$
($c^S_j$, respectively) the {\it starting time} ({\it completion
time}, respectively) of job $j$ in a schedule $S$. A {\it feasible
schedule} is a schedule which satisfies all restrictions of the
given problem. Since all problems imply resource restrictions, in
any feasible schedule, if job $j$ is scheduled on the machine, then
$s_j^S\ge s_i^S+p_i$, for any job $i$ scheduled earlier on the
machine in $S$ (recal that the machine can handle only one job at any
time). Besides, if $j$ has a release time then it cannot start
before this time, i.e., $s_j^S\ge r_j$; note that $c^S_j =
s^S_j+p_j$ whenever no job preemption is allowed. An {\it optimal
schedule} is a one which minimizes the given objective (goal)
function.

Graham's three-field notation is a common way to 
abbreviate a  scheduling problem. The first, the
second and the third fields define the machine environment, the
job characteristics and the objective function, respectively. In a
{\it single-machine (single-processor) scheduling problem}, 
a ``1'' is used in the first field in the three-field
notation to specify a single machine environment. For example, 
$1|r_j,q_j|C_{\max}$ represents the single machine scheduling
problem with release times and tails to minimize the maximum full  
job completion time (the makespan). 

 There exist a huge number of papers dealing with single machine problems. In particular, many heuristics and metaheuristic as well as enumerative algorithms have been suggested and tested for special single machine problems. Regarding surveys, the authors often restrict to special subjects and/or consider more general problem settings. As an example, there exist three very detailed surveys on scheduling problems with setup times or costs, see \cite{allah1, allah2, allah3}. For  problems with so-called $p$-batching (i.e., several jobs can be processed in parallel in a batch), a survey has been given by Fowler and M\"onch \cite{fm}. The case of equal processing times has been surveyed by Kravchenko and Werner \cite{kw}. 

Regarding single machine problems, often surveys are given for particular objective functions. As a few examples, we can mention e.g. a survey on the total weighted tardiness problem \cite{apw}, on the weighted number of tardy jobs problem \cite{aa} or again for more general problems, a survey on scheduling with late work criteria \cite{sterna1}, or more recently on early and late work scheduling \cite{sterna2}, where a substantial part is dedicated to single machine scheduling.
While mostly a minimization problem is considered, Gafarov et al. \cite{GLW13} reviewed algorithms and complexity issues for single machine maximization problems.

Finally we note that single machine problems are also important for some practical multi-machine scheduling applications. Often there exists a bottleneck machine in a manufacturing environment. If one can solve the scheduling problem on the bottleneck machine optimally, this is at least a very good base for the overall schedule.

In this paper, we review some results related to polynomial algorithms, the complexity and  approximability issues. We explicitly mention that we do not survey metaheuristic and enumerative algorithms. In addition to some necessary basic results, the main focus is on the results which have been presented in the past in papers, where the authors of this work participated.

The rest of this paper is organized as follows. In Section 2, we review some necessary basics for single machine scheduling problems. Then, in Section 3, 
we discuss polynomial algorithms, complexity and approximability issues for different classes of single machine scheduling problems. We also outline some subjects for future research, partially also for more general multi-machine scheduling problems. 
The paper ends with some concluding remarks in Section 4.   

\section{Some Basics}

In this section, we give some basic notations, concepts and properties 
that have been used in the study of single machine scheduling problems. 
Single machine scheduling problems are trivial if every job has a
single parameter, its processing time. For minimizing the makespan, 
it suffices to construct any feasible schedule without a machine idle
time or a gap including jobs in any order: 

A {\it gap}  on a machine is a longest time interval during which
this machine is idle, i.e., it processes no job.  It can be assumed that 
there occurs a 0-length gap $(c_j,t_i)$ if job $i$ starts at
its release time at the the completion time of job $j$.

A {\it block} is a consecutive part in a schedule  which is
preceded and succeeded by a gap.

Single machine scheduling problems get less trivial if
jobs are not simultaneously released, i.e., they have also different
release times. Then, to avoid the creation of any redundant gap, we
just need to keep the machine busy whenever this is possible. It is
easy to see that this goal will be achieved if we just order the jobs 
according to non-decreasing release times. After this preprocessing step, the jobs
can be included in this order, again, leaving no avoidable gap. 
This procedure will take $O(n\log n)$ time due to the preprocessing
step. 

If both, job release and due dates are given, then it makes sense
to minimize the maximum job lateness (note that due dates are irrelevant 
for the makespan minimization). Suppose preemptions are allowed. 
Then we have a simple method to solve the problem optimally even
on-line. We usually start at the earliest job release time and
set initially the current time $t$ to that magnitude. Iteratively, 
among the available jobs, we schedule the next a most urgent job, 
one with the smallest due date. However, if during the execution of
that job, another more urgent job gets released, the former job 
is interrupted at the release time of the latter job, which gets
included at its release time. In this way, there will occur no forced
delay for any urgent job by a non-urgent one, unlike the case
without preemptions (in a given schedule $S$, the {\em delay} of 
job $j$ is $t_j^S - r_j$). The so constructed schedule is compact 
also in the sense that it minimizes the makespan. In particular, 
it has no avoidable machine idle time.

Assume that job $i$ is processed at time $r_j$ in the schedule $S$, where $d_i > d_j$.
Without allowing the interruption of $i$, this job delays the starting of job
$j$; note that this delay will be less than $p_i$ since upon the completion 
of job $i$, job $j$ can immediately be scheduled. The delay of job $j$ may 
result in a non-optimal schedule. Indeed, this will be the case 
when job $j$ is too urgent with  a ``small'' due-date. In an optimal
schedule, it should be included with no or with a smaller delay. 

Now we describe a basic popular greedy non-preemptive algorithm 
to which we will refer as {\it earliest due-date} (ED) heuristic). 
It that has been widely used for solution of different single
machine and other more complex scheduling problems. As the above 
described preemptive one, it is on-line. The heuristic was suggested by 
Jackson \cite{J} and later developed by Schrage \cite{Sch}. 
The first version is just the above described greedy algorithm with the
pre-processing stage. The second version was adopted for jobs having both, 
release and due times. We describe this heuristic now. 
We let the current scheduling time $t$ be the minimal release time of 
an unscheduled job or the time when the machine completes the last
job, whichever magnitude is larger. If no job is yet scheduled, 
we let $t:=0$. Iteratively, among the jobs already released by time 
$t$, the heuristic schedules one with the smallest due date. It breaks
 ties selecting a longest job. 

Let $\sigma$ be the {\it initial ED-schedule}  generated by the 
ED-heuristic for the originally given problem instance. Since no urgent
job can be delayed by more than the processing time of another job, it
is easy to see that $$L_{\max}(\sigma) < p_{\max}=\max_{j\in J}p_j.$$
In particular, $\sigma$ is not necessarily optimal. 

Let us introduce some additional concepts that will be useful to understand
better the problem. Let $S$ be an ED-schedule, one generated by the ED-heuristic.

Among all jobs in the schedule $S$, it is useful to distinguish the ones whose 
lateness attains $L_{\max}(S)$. In a given block $B\in S$, let us call among them the last 
scheduled job from this block $B$ an {\it overflow job} in the schedule $S$. Let 
$o(S)$ be the earliest overflow job in the schedule $S$, and let 
the {\it critical block} $B(S)$ of $S$ be the block containing job 
$o(S)$. 

Let now  $e$ be a job from the block $B(S)$ such that  $d_e > d_{o(S)}$, i.e., it
is less urgent than the overflow job $o(S)$. Then job $e$ is referred to as
an {\it emerging job} in the schedule $S$. The latest scheduled emerging job 
in the block $B(S)$ is called the {\it delaying} emerging job.

The {\it kernel} $K(S)$ in the schedule $S$  consists of the set of jobs,
included in that schedule between the delaying emerging job $l(S)$ and the
overflow job $o(S)$, not including job $l(S)$ (but including job $o(S)$). 
It immediately follows that the due date of any job of the kernel $K(S)$ is no 
less than that of job $o(S)$). 

The above definitions can straightforwardly be applied to any (not necessarily
critical) block. Then there may exist more than one overflow job (at most one in
in each block) in a schedule. Note that we will have the same number of kernels 
in that schedule. 
 
The {\it earliest starting time} of a kernel $K$ is given by  $${\cal S}(K)=\min\{r_j|j\in K\}.$$
The {\it starting time} ${\cal S}(K,S)$ of kernel  $K$ in the schedule $S$
 is the starting time of the first scheduled
job of that kernel in the schedule $S$.

If the ED-schedule $S$ is not optimal, the overflow job $o(S)$ is to be restarted earlier. 
By modifying artificially the initial problem parameters and applying the ED-heuristic 
to this modified instance, an alternative ED-schedule in which job $o(S)$ is 
restarted earlier can be created, as we describe now. In the schedule $S$, we look 
for the overflow job $o(S)$, the kernel $K(S)$ and the set of all emerging jobs. 
In an attempt to reduce $L_{\max}(S)$ (restart job $o(S)$ earlier), 
we {\it activate} an emerging job $l$ for the 
kernel $K(S)$, that is, we reschedule job $l$ after the kernel $K(S)$. 
The activation is carried out in two steps.  At step 1  the
release time of job $l$ is increased artificially to it a magnitude, no-less than the
largest job release time in the kernel $K(S)$. At step, ED-heuristic is applied
to the modified in this way problem instance. Since now the
release time of job $l$ is no less than that of any job of the kernel $K(S)$, and
$d_l$ is larger than the due date of any job from the kernel $K(S)$, the ED-heuristic 
will give the  priority to the jobs of the kernel $K(S)$ and will reschedule job
$l$ after all former jobs. The ED-schedule  $\{S\}_l$, obtained in this way, 
is called a {\it (weakly) complementary} to $S$ schedule.

Note that in the schedule $S$, some emerging jobs might have been originally 
included after the kernel $K(S)$. Then at least one such job will be included before 
the jobs of the kernel $K(S)$ in the schedule $\{S\}_l$ by the ED-heuristic (since no job of 
the kernel $K(S)$ is released at time $s_l(S)$). As a result, the kernel $K(S)$ may not be 
restarted at its earliest starting time. To guarantee
the restarting of the kernel $K(S)$ at its earliest starting time, the above
emerging jobs might be forced to remain after the kernel $K(S)$. To reach this
goal, we increase the release times of these jobs, 
similarly as we did it for  job $l$. The ED-schedule, obtained by the ED-heuristic 
for the modified in this way problem instance, is called  a {\it (strongly)
complementary} to $S$ schedule and is denoted by $S_l$. A complementary schedule 
can be abbreviated as a {\it C-schedule}.

Note that there is exactly one job less included before the kernel $K(S)$ in the 
schedule $S_l$ compared to schedule $S$. (The same condition does not held in
$\{S\}_l$; in fact, we may have more jobs scheduled before the kernel $K(S)$
in the schedule $\{S\}_l$ than in the schedule $S$.) Then, as it is easy to see, 
the overflow job $o(S)$ will be completed earlier in the schedule $S_l$ than it was 
completed in the schedule $S$:

\begin{lemma}\label{gap} Let $S$ be  a strongly C-schedule and let 
$l$ be the emerging job scheduled in $i$th position in in that schedule.
Let, further $j$ be the job, immediately succeeding job $l$ in 
schedule $S$. Then if $d_j < d_l$, there will occur a gap in between 
the $i$th scheduled job and the kernel $K(S)$ in the schedule $S_l$ . 
Furthermore, there will arise a gap immediately before the earliest
scheduled job of the kernel $K(S)$ in the complementary schedule $S_l$, 
where $l$ is the live emerging job in the schedule $S$. As a result, 
the kernel $K(S)$ will start at the minimum release time of a job from
that kernel.
\end{lemma}

Proof. Assume first that job $j$ is scheduled in the $i$th position
in the schedule $S_l$ and there arises no gap immediately before that
job in that schedule. Then $r_j \le r_l$ holds, but since $d_j < d_l$,
the ED-heuristic would schedule job $j$ instead of job $l$ at
the moment $t_l(S)$ in $S$, which is a contradiction. We
use a similar reasoning if instead of job $j$ another
job $k$ with $d_k < d_l$ is scheduled in the $i$th position
in schedule $S_l$. If $d_k \ge d_l$, 
since the schedule $S_l$ is a strongly C-schedule, job $k$ must be an
emerging job scheduled between job $l$ and kernel $K(S)$ in $S$.
However, there is scheduled one job less before the kernel $K(S)$ in the  
schedule $S_l$ than in the schedule $S$, and our claim easily follows.\eop  

To distinguish the type of a gap in Lemma \ref{gap} from  a  ``natural'' 
gap that arises in the initial ED-schedule, the former type of a gap is 
sometimes referred to as an ``artificial'' one.

\section{Polynomially Solvable Cases, Complexity and Approximability Issues}

As earlier noted, the non-preemptive single machine scheduling problem becomes 
non-trivial, if, besides job the processing times, a job has at least one 
more parameter (as otherwise, from {\em any} permutation of the $n$ jobs, 
an optimal makespan schedule can be constructed in linear time
by assigning the jobs from that permutation without leaving any idle time
interval on the machine). In particular, all jobs are simultaneously released, 
any non-idle time schedule minimizes the makespan $C_{\max}$.

Minimizing maximum job lateness $L_{\max}$ is a bit more complex (recall 
that the lateness of job $j$ in a schedule $S$ is $L_j^S = c^S_j-d_j$).  
If only the job due dates are given (all the jobs being released at time 0), 
the Earliest Due date (ED) heuristic obtains an optimal schedule 
minimizing  maximum job lateness (Jackson \cite{J}). 
This heuristic schedules the jobs 
in non-decreasing order of their due dates without any idle time
intervals (gaps). Vice versa, if only job release times are given, 
the Earliest Release time (ER) heuristic generates an optimal makespan 
schedule. In that heuristic, initially, the minimum job release time 
is assigned to the current scheduling time, and any job released at
that time is assigned to the machine. Iteratively, the next scheduling
time is defined as the maximum between the completion time of the
latest so far assigned job and the minimum release time taken among the 
yet unscheduled jobs. Again, any (unscheduled) job released at the 
current scheduling time is assigned to the machine. 

Because of the job release times, a schedule created by the ER-heuristic may 
contain a gap. However, it may contain no gap that might be avoided. In 
particular, the total gap length in such a schedule is the minimal
possible one. We can also use this  heuristic if all 
the jobs have a common due date $d$. In other words, all jobs are
equally urgent. Then, as it is easy to see,  an optimal schedule 
minimizing the maximum job lateness contains no gap that might be
avoided, and it has the minimum possible total gap length. Therefore, 
the ER-heuristic minimizes also the maximum job lateness. 

It is easy to see that the ER-heuristic will no more minimize the 
maximum job lateness even if only two possible job due dates are 
allowed. Indeed, consider an instance with two jobs, job 1 
released at time 0 and having the due date 11 and job 2 released
at time 1 with the due date 6. The ER-heuristic will assign job 1
to the time interval $[0,5)$ and job 2 to the time interval 
$[5,10)$ resulting in the maximum job lateness 4 (that of job 2).
Observe that this schedule contains no gap. At the same time, an 
optimal schedule does contain a gap $[0,1)$: it assigns job 2 to 
the time interval $[1,6)$ and then job 1 to the time interval 
$[6,11)$ completing both jobs on-time with the maximum job lateness 
equal to 0.

\subsection{Scheduling a Single Machine with Release Times and Due Dates
to Minimize Maximum Job Lateness}

The general problem with release tames and due dates to 
minimize maximum lateness, abbreviated as $1|r_j|L_{\max}$,
is strongly NP-hard  \cite{GJ}. If we replace the due dates by tails
with the objective to minimize maximum lateness, we get
an equivalent problem $1|r_j,q_j|C_{\max}$. In this transformation, 
every due date $d_j$ is replaced by the tail $q_j=A-d_j$, where $A$
is a sufficiently large constant (no less than the maximum job due date). 
So, the smaller is the due date, the larger is the tail. In other words, 
the urgency of every job is conserved with this transformation. 
It is straightforward to see that the two versions are
equivalent. The {\it full completion time} of job $j$ in $S$ is given by ${\cal
C}_j(S)=c_j^S+q_j$. The objective in that setting  is to find an optimal 
makespan schedule, i.e., a feasible schedule $S$ with the minimal
value of the maximal {\em full} job completion time.

An efficient implicit enumerative algorithms for the problems
$1|r_j|L_{\max}$ and $1|r_j,q_j|C_{\max}$ was 
suggested in McMahon and Florian \cite{Mc} and later, based on similar ideas, in  Carlier \cite{C}. 
A number of other special cases are solvable in polynomial 
time. If all jobs are of unit length,  
the problem is again easy to solve, see Horn \cite{horn}: indeed, while applying the 
ED-heuristic, no new (earlier unreleased) job can be released within the 
execution interval of a currently running job since job release times are 
integer and every job takes just a unit time of processing. As a result, 
no job may cause a forced delay of any more urgent job. 

As we have already seen, if all $r_j$s are equal, then
scheduling the jobs in order of non-decreasing due dates gives an
optimal schedule in $O(n\log n)$ time \cite{J}. Similarly, if all
$d_j$s are equal, then scheduling the jobs in order of non-decreasing
release times is optimal. An $O(n\log n)$ solution of the problem was 
proposed in \cite{H} for the special case when the release times, 
processing times and due-dates are restricted in a specific manner 
so that each $r_j$ lies within a certain time interval defined 
by $p_j$ and $d_j$.

With arbitrary release times and due dates (or tails), if all processing 
times are equal (but not necessarily of unit length), it is still possible 
to find an optimal schedule in polynomial time, see Garey et al. 
\cite{GJS}. Even with 
two allowable job processing times $p$ and $2p$, for an arbitrary
integer $p$, the problem 
$1|p_j\in \{ p,2p\},r_j,q_j|C_{\max}$ remains polynomially solvable \cite{aor04}. 
In fact, in the algorithm from the latter reference the processing times 
of only some jobs (which become known during the execution of the 
algorithm) are restricted to $p$ or $2p$, whereas the rest of the 
jobs may have arbitrary processing times. The algorithm
is obtained as a consequence of two enumerative algorithms for
the problem $1|r_j,q_j|C_{\max}$. The first algorithm is exponential. 
The the second algorithm is a restriction of the first one and runs in
time $O(n^2\log n)$  under certain conditions which are to be held during 
its execution. Otherwise, another auxiliary procedure
is to be applied to guarantee the optimality. An $O(n^2\log n\log P)$ 
implementation of that procedure for the problem  
$1|p_j\in \{ P,2P\},r_j,q_j|C_{\max}$ yields
an algorithm with the same time complexity for this setting.

If the maximum job processing time $p_{\max}$ is bounded from above by 
a polynomial function in $n$ (i.e., $P(n)=O(n^k)$) and the maximum difference between 
the job release times is bounded by a constant $R$, then the corresponding
parameterized  problem $1| p_{max} \le P(n), |r_j-r_i|<R \ | L_{\max}$ can  
still be solved in polynomial time $O(n^{k+1}\log n \log p_{\max})$, see 
Vakhania \& Werner \cite{tcs13}. Note that in this setting the general 
problem is parameterized by $p_{\max}$ and the maximum difference between 
two job release times. 

There are known other conditions that allow a polynomial time solution of
the problem. For example, if the job parameters are embedded in a special manner, 
i.e., for any pair of jobs $j,i$ with $r_i > r_j$ and 
$d_i < d_j$, $$d_j-r_j-p_j\le d_i-r_i-p_i$$ holds, and with $$r_i+p_i \ge r_j+p_j,$$
$d_i \ge d_j$ holds, then the resultant problem  
$1|r_j,embedded|L_{\max}$ is solvable in $O(n\log n)$ time, see Vakhania \cite{tcs19}. 

In a maximal polynomially solvable special case of the problem  
$1|p_j:divisible,r_j|L_{\max}$, the job processing times form mutually divisible numbers 
(say, powers of 2)  \cite{1div}. 

{For the problem $1|r_j, p_j = p|L_{max}$ with constant processing times, Lazarev et al. \cite{LAW} presented two polynomial
	algorithms. The first procedure is based on trisection search and determines an optimal solution in $O(Q n \log n)$ time, where 
	$10^{-Q}$ represents the accuracy of the parameters of the problem. The second approach uses an auxiliary problem, where the maximum completion time is minimized under the constraint that 
	$L_{max}$ is bounded  by a particular value. Using this approach, the authors determine for the bi-criteria problem with the two criteria $C_{max}$ and $L_{max}$ and the Pareto set 	of optimal solutions in $O(n^3 \log n)$ time.

With arbitrary job processing times, the problem remains (weakly) $NP$-hard 
already with only two allowable job release times and due dates
$1|r_j\in \{r_1,r_2\},d_j\in \{d_1,d_2\}|L_{\max}$, see Chinos and Vakhania 
\cite{elisa}. This is a minimal $NP$-hard special case of the problem 
$1|r_j,d_j|L_{\max}$.  The latter setting was later studied in 
Reynoso and Vakhania \cite{alej}, where structural properties of this 
special case have led to a number of optimality conditions under 
which it can be solved optimally in $O(n \log n)$ time were established. 
It was shown that, for bottleneck instances where none of these 
conditions apply, the problem can be reduced to the SUBSET SUM problem. 
The authors presented computational results accomplished for 50 
million randomly generated problem instances, reporting  that
almost all these instances were solved optimally by one of the
established optimality conditions. Moreover, the heuristic algorithms
that incorporate the optimality conditions established in \cite{alej} 
succeeded to solve the SUBSET SUM problem also optimally for almost all 
the 50 million instances. A very recent work considered a natural 
generalization of the latter problem with any fixed number of 
release times, see Escalona and Vakhania \cite{diana}. 
For this setting, the optimality conditions were derived and again, 
it was shown how the problem can be reduced to the SUBSET SUM. It is
a challenging question to what extent the above obtained results can be 
generalized for any fixed number of release and delivery times, a setting
with much wider practical applications.

In \cite{AGW} a more general objective function $f_{max}$ defined as the maximum of non-decreasing cost functions $f_j$ has been considered for the preemptive problem $1|r_j,prec,pmtn|f_{max}$. It is assumed that the cost functions fulfill the order relation $f_i(t) \leq f_j(t)$ or $f_j(t) \leq f_i(t)$ for any pair $i, j$ of jobs and for any time $t$ of the scheduling interval. In the case of tree-like precedence constraints, the derived algorithm has a time complexity of $O(n \log n)$. In addition, a parallel algorithm is given with a time complexity of $O(\log^2 n)$ when using $O(n^2 \log n)$ processors.   

A special kind of parameterized analysis called the
{\em Variable Parameter} (VP)  analysis was recently
suggested in \cite{compact}. Based on this approach,  an implicit enumeration algorithm and a polynomial time approximation scheme
for the general setting 
$1|r_j, q_j|C_{\max}$ were proposed. The variable parameter is 
defined as the number of emerging jobs. 
First, a partial solution in time $O(n\log n)$ without these jobs
is constructed. Then this partial solution is  augmented to a complete optimal solution in time, which is exponential solely in the number 
of the variable parameter (the number of emerging jobs).  
The author also gives alternative time complexity expression for
both, the implicit enumeration and the polynomial time approximation scheme, in which the exponential dependence is solely on the maximum
processing time and the maximum delivery time of an emerging job. 
If one applies the fixed parameter analysis to these estimations,
one gets a polynomial-time dependence.  A very recent extensive experimental study showed that, in practice,  
the variable parameter is much  less than $n$. Moreover, the ratio 
of the variable parameter to $n$ asymptotically converges to 0
(see  Ramirez and Vakhania \cite{kevin}).

The problem $1|r_j|L_{\max}$ is important on its own and also because
of its possible applications for the solution of other more complex scheduling
problems and other optimization problems. For example, in the Multiple Traveling
Salesman Problem (M-TSP), $k$ salesmen, each starting from a special point $0$ called
depot, has to visit a number of clients and return to the depot. The objective
is to minimize the total cost of all $k$ tours (see, for example 
\cite{TSP axioms}). In the setting with {\em time windows} M-TSP-TW, each client 
has to be visited within a specified time interval. In the former basic setting, 
it is easy to find a feasible solution to the problem. However, in the latter
setting with time windows, there may exist no feasible solution for a given 
$k$. Hence, it is crucial to determine whether there exists a feasible solution 
for a given $k$. More generally, one may be interested to calculate the
minimum $k$ for which  there exists a feasible solution. We can use our scheduling
problem for this purpose. 

In the basic case, we wish to determine whether there exists 
a feasible solution with $k=1$. For a given instance of  M-TSP-TW with $n+1$
clients, we create an instance of he problem $1|r_j, p_j=0|L_{\max}$ with $n$ jobs, associating
job $j$ with client $j$, for each $i=1,\dots,n$, and letting the processing
time of each job to be 0. Note that this is a simplified version of  the 
generic problem $1|r_j|L_{\max}$.  In this version, the release and due 
times of each job are determined
according to the time window of the corresponding client. We consider the 
feasibility version of the problem, where one looks for a schedule in which 
no job has a positive lateness. One can immediately see that, if there is no 
feasible solution to the scheduling problem, there is also no feasible
solution for the corresponding instance of the problem M-TSP-TW with $k=1$ (i.e., TSP-TW). 
In general, we let $u$ be an upper bound on the total number of salesmen, and 
carry out binary search within the interval $[1,u]$ solving at each 
iteration an instance of the corresponding scheduling problem 
$Pk|r_j, p_j=0|L_{\max}$ with  $k$ parallel identical machines and with
trial $k\in [1,u]$. In this way, we can find the minimum
$k$ for which there exists a feasible solution to the problem M-TSP-TW.

\subsection{Minimizing the Number of Late Jobs}

Another common objective function with given due dates is the {\em number
of late jobs}: a job in a schedule is called {\em late} if it is
completed after its due date in that schedule, otherwise it is called {\em on time}. 
In the problem of this subsection, we deal  with $n$ jobs with release times 
and due dates which have to be scheduled on the machine so as to 
maximize the number of on time jobs, or equivalently,  to
minimize the number of late jobs. We consider both, preemptive and non-preemptive
cases, and also cases when the jobs have an additional parameter called {\em weight}.

\subsubsection{Nonpreemptive Scheduling to Minimize the Number of
Late Jobs: $1|r_j;p_j=p|\sum U_j$}

In this subsection, we deal with the non-preemptive version of the problem
$1|r_j|\sum U_j$. This problem was studied in Vakhania \cite{jos13}, 
where two polynomial time algorithms were proposed. The first one
with the time complexity $O(n^3 \log n)$ solves optimally the problem 
if during its execution no job with some specially defined property occurs. 
The second algorithm is an adaptation of the first one for the special case 
of the problem when 
all jobs have the same length $1|r_j;p_j=p|\sum U_j$. The time complexity 
of this algorithm is $O(n^3 \log n)$ (it improved an earlier known fastest algorithm with the time complexity
of $O(n^5)$  by Chrobak et al.  \cite{Chrob}.

\subsubsection{Preemptive Scheduling to Maximize the Number of
on Time Jobs: $1|r_j;pmtn|\sum U_j$}

In the problem of this subsection, $n$ jobs
with release times and due dates have to be scheduled preemptively
on a single machine so as to minimize the number of late jobs, i.e., problem
$1|r_j;pmtn|\sum U_j$. We call a job {\it late} ({\it on time},
respectively) if it is completed after (at or before,
respectively) its due date. We will say that job $j$ is {\it
feasibly} scheduled if it starts no earlier than at time $r_j$ and
overlaps with no other job on the machine. A feasible schedule is
one which feasibly schedules all jobs. The objective is to find a
feasible schedule with a minimal number of late jobs,
or equivalently, with the maximal number of on time jobs.

This problem is known to be polynomially solvable. The first polynomial
dynamic programming algorithm with a time complexity of $O(n^5)$
and a space complexity of $O(n^3)$ was developed by Lawler
\cite{law1}. The result was improved by Baptiste \cite{bapt} who
suggested another dynamic programming algorithm with a time
complexity of $O(n^4)$ and a space complexity of $O(n^2)$. 
The latter result was further improved by Vakhania \cite{orl09},
where an algorithm with the time and space complexities of $O(n^3)$ and
$O(n)$, respectively, was proposed. Unlike the earlier algorithms, the latter
algorithm is not based on dynamic programming. It uses the ED-heuristic 
for scheduling the jobs. The main task is to
construct a schedule in which all jobs are scheduled on time,
with the number of such jobs being maximal. As any late job can be
scheduled arbitrarily late without affecting the objective
function, jobs not included in the former schedule can be later added at the end
of it in an arbitrary order. During the construction of the schedule, some
jobs are finally disregarded, whereas some other jobs are
omitted with the possibility to be included into a later created schedule.

\subsection{Preemptive Scheduling to Minimize the Weighted Number of
Late Jobs with Equal Length: $1|r_j; p_j = p; pmtn|\sum w_j
U_j$}

In this section, we consider the weighted version of the problem of the
previous section in which preemptions are allowed and all jobs
have an equal length. Moreover, $w_j$ is the non-negative integer weight
of job $j$. We minimize the {\it weighted throughput} which is the
total weight of the completed jobs. Equivalently, we minimize the
weighted number of late jobs. The problem ican be abbreviated as
$1|r_j; p_j = p; pmtn|\sum w_j U_j$.

The first polynomial dynamic programming algorithm of the complexity $O(n^{10})$ for
the problem was suggested by Baptiste~\cite{baptiste99a}. Baptiste et al. 
\cite{orl04-2}
improved the above time complexity and proposed another dynamic
programming algorithm which runs in $O(n^4)$ time.

Next, we briefly discuss some directions for further research.
First we note that for the multi-processor
case, the weighted version is $NP$-complete \cite{BK99}. At the same time, 
the non-weighted version still remains open. In particular, it
is not known whether the problem $P|r_j; p_j = p; {pmtn}|\sum
U_j$ can be solved in polynomial time. (Even for $2$ processors, 
one cannot the search to solely earliest deadline schedules. 
As an example, consider an instance   
with three jobs with feasible intervals $(0,3)$, $(0,4)$,
and $(0,5)$ and the processing time $p=3$. There is a feasible
schedule that completes all three jobs, whereas the
earliest-deadline schedule completes only jobs $1$ and $2$.) 
For the multi-processor case, it is also interesting to study the preemptive
version, where jobs are not allowed to migrate between processors.

\subsection{Scheduling to Minimize the Weighted Number of Late jobs with Deadlines and  Release/Due Date Intervals}

 In \cite{GPW}, Gordon et al. dealt with the problem, where in addition to release and due dates also deadlines are given. This means that certain jobs have to be completed on time. The authors considered the two cases of similarly ordered release and dues dates (i.e., $r_1 \leq r_2 \leq \ldots \leq r_n$ and $d_1 \leq d_2 \leq \ldots \leq d_n$ as well nested release and due date intervals (i.e., none of two such intervals do overlap, this they have at most joint point or one covers the other one). For these problems, a reduced problem is constructed, where the deadlines are eliminated, and it is shown that the optimal solution of the reduced problem can be used to find an optimal solution of the initial problem. 

In a later paper, Gordon et al. \cite{GWY} considered the preemptive version of the above problem with nested release and due date intervals. They derived necessary and sufficient conditions for the existence of a schedule in which all jobs are completed by their due dates. For the case of oppositely ordered processing times and job weights, a polynomial algorithm of the complexity $O(n \log n)$ has been presented.       

\subsection{Scheduling with Tardiness-based Objectives}

While the problem $1||\sum T_j$ is known to be NP-hard, Gafarov et al. \cite{GLW20} considered the special case of equal processing times, i.e., problem $1|r_j, p_j = p|\sum w_jT_j$. After deriving some properties, they gave polynomial algorithms for two special cases of the problem. If the difference between the maximum due date and the minimum due date is not greater than the processing time $p$, the problem can be solved by a modification of the algorithm by Baptiste \cite{bap} in $O(n^{10})$ time.
The special case with ordered weights and processing times according to $w_1 \leq w_2 \leq \ldots \leq w_n$ and $d_1 \leq d_2 \leq \ldots \leq d_n$ can be solved in $O(n^9)$ time by a similar modification from \cite{bap}.     

In \cite{LW1}, Lazarev and Werner presented a new general graphical algorithm which often improves the complexity or at least the running time of dynamic programming algorithms, and it can be used for efficient approximation algorithm. Compared to standard dynamic programming algorithms, the number of states can often substantially be reduced. 

In a standard dynamic programming algorithm for permutation problems, functions $F_l(t)$ and the corresponding partial sequences have to be stored for any stage $l = 1, 2, \ldots, n$ and any time $t = 0, 1, \ldots, UB$.  Here $UB$ is an upper bound for the scheduling interval, e.g. the sum of all processing times (or the maximal due date). The values $F_1(t)$ typically represent the best function value when considering the first $l$ jobs and allowing at most $t$ time units for the processing of the scheduled (e.g. early) jobs. This results in a pseudo-polynomial algorithm of the complexity $O(n \sum p_j)$ . 

The graphical algorithm combines different states, where the best partial solution does not change and the corresponding objective function $F_l(t)$ can be described by the same linear function. As a result, at each stage $l$, there are certain intervals separated by the so-called breakpoints $t^l_1 < t^l_2 < \ldots t^l_{h_l}$, where the structure of the linear function belonging to the best partial solution changes. In addition, for the graphical algorithm, the times $t$ do not need to be integer.

In \cite{GDW}, Gafarov et al. described the application of this graphical algorithm to the approximate solution of various single machine  problems. In particular, the graphical algorithm is applied to 5 total tardiness problems, where a pseudo-polynomial algorithm exists, and the graphical algorithm yields fully polynomial-time approximation schemes yielding the best known running time.

In \cite{GLW13}, Gafarov et al. considered the maximization of total tardiness and by applying the graphical algorithm, they transformed a pseudo-polynomial algorithm into a polynomial one. More precisely, the complexity of $O(n \sum p_j)$ can be reduced to $O(n^2)$ what settled the open complexity status of the problem $\max 1|(nd)|\sum T_j$),  where $nd$ means that no inserted idle times occur. The authors present a detailed numerical example and give also a practical application of such maximization problems, where wind turbines have to be mounted in several regions of a country.

While for more general settings, such a graphical algorithm improves often the running time in comparison with existing algorithms, the complexity remains pseudo-polynomial. For instance, in \cite{GLW12}, Lazarev et al. presented graphical algorithms with a complexity of $O(n \sum p_j)$ for a special case of a generalized tardiness problem corresponding to the minimization of late work as well as for the minimization of the weighted number of late jobs.

In \cite{LW2}, Lazarev and Werner considered various special cases of the (unweighted) total tardiness problem $1||\sum T_j$. After proving that already the special case of oppositely ordered processing times and due dates, i.e. $p_1 \geq p_2 \geq \ldots \geq p_n$ and $d_1 \leq d_2 \leq \ldots \leq d_n$, is $NP$-hard in the ordinary sense, they gave several pseudo-polynomial and polynomial algorithms for special cases of this problem. In particular, $O(n^2)$ algorithms have been given for the cases $d_j - d_{j-1} > p_j, j = 2, 3, \ldots, n$ as well as for the case that the difference between the maximal and the minimal due date is not greater than 1. In both cases, the processing times do not need to be integer.

In further research, the graphical algorithm can be applied to more general problems, where a pseudo-polynomial algorithm exists and Boolean variables are used for making yes/no decisions.

\subsection{Preemptive Scheduling in Overloaded Systems}

In the single machine scheduling problem in overloaded systems, 
each job has a release time, a deadline and a weight. The latter
parameter reflects its profit. Any job can be preempted during 
the execution, and the objective is to maximize the overall profit. 
Two possibly models and the definition of the profit gained 
by the execution of task $j$ can be defined.  In
the \emph{standard model}, a completed task $j$ gives the profit
$w_jp_j$, whereas a non-completed task gives no profit. In the
\emph{metered model}, a task $j$ executed for $t\leq p_j$ time
units gives the profit $w_{jt}$ (this task may not be  completed).
For the off-line metered profit model, Chrobak et al. \cite{jcss03}
used a reduction to bipartite matchings and maximal
flows to develop a polynomial-time algorithm. 
For the online metered case, they presented an
algorithm with competitive ratio $e/(e-1)\approx 1.5820$ and 
proved a lower lower bound of $\sqrt{5}-1\approx 1.236$ on the
competitive ratio of algorithms for this problem.

\subsection{Scheduling with Restarts}

In the on-line single machine scheduling problem with restarts, jobs arriving
over time are characterized by processing, release and delivery times, 
which become known at the arrival (release) time of each job. The objective 
is to minimize the maximum full job completion time, the makespan. 
As earlier mentioned, the corresponding off-line problem is strongly
NP-hard, but the preemptive version can be solved in polynomial
time by the preemptive LDT-heuristic. For the non-preemptive case,
the LDT-heuristic has a performance guarantee of 2, which is the best
possible one unless we allow the machine to be idle when there is an
unprocessed job available. Hoogeveen and Vestjens \cite{hv00}
developed an on-line algorithm with the best possible worst-case ratio 
$$(\sqrt{5}+1)/2 \approx 1.61803$$ in which a waiting strategy is incorporated. 
Van den Akker, Hoogeveen and Vakhania \cite{jos00}  considered
an extension of the standard model by allowing job {\em restarts}: if
an urgent job comes in while the machine is busy, then  the currently 
running job can be interrupted in favor of the former urgent job. In 
this model, unlike traditional preemptive models, the already processed part of the
interrupted job is wasted, this job is to be started all over again. The
algorithm proposed in \cite{jos00} has the best possible worst-case ratio of 
$1.5$ (which is better than the above mentioned bound for the case without
the restarts).

Again, we give some possible directions for further research in connection 
with allowed restarts. As we just mentioned, with the restart model 
considered in \cite{jos00}, if a job is restarted, then the portion of
this job that was already processed by the machine is lost. This may be
avoided for some applications, where the already completed part of an 
abandoned job gives some benefit and this job can be restarted from the
time moment when it was interrupted or at some earlier time. Hence, one
may consider models with partial restarts when the already processed
part of an interrupted job counts partly or completely. In particular, 
it is interesting to find out whether there exist better worst-case 
performance bounds for models with partial restarts.

\subsection{Scheduling with Financial Resource Constraints}

Gafarov et al. \cite{GLW0} considered single machine problems with a non-renewable resource (abbreviated $NR$) like e.g. money or fuel. This means that at any time $t_i, i = 0, 1, \ldots, UB$ with $t_0 = 0$ and $UB$ as an upper bound on the scheduling interval, one gets an amount $G(t_i) \geq 0$ of the resource, while any job $j$ has a consumption of $g_j$ at the starting time $s_j$ of the job. Hence, the equality
$$  \sum_{j=1}^n g_j = \sum_{i=0}^{UB} G(t_i) $$
holds. One looks for an optimal permutation $(j_1, j_2, \ldots, j_n)$ for a particular objective function with the starting times $s_{j_1}, s_{j_2}, \ldots, s_{j_n}$ of the jobs such that the resource constraints are satisfied, i.e., the inequality  
$$   \sum_{l=1}^i g_{j_l} \leq   \sum \{ G(t_v) \mid t_v \leq s_{j_i} \mbox{ for all } v \leq s_{j_i} \}   $$
holds for any $i = 1, 2, \ldots, n$.
Such problems play a role e.g. for government-financed organizations when a project can only be started after receiving the required funds. 

The authors focused on complexity results and proved that the problems $1|NR|f$ with $f \in  \{ C_{max},\sum C_j, \sum U_j\}$ are strongly $NP$-hard. When minimizing total tardiness $\sum T_j$, the case of equal due dates ($d_j = d$) is already strongly $NP$-hard. The following special cases with minimizing total tardiness were proven to be weakly $NP$-hard: \\[1ex]
1) equal processing times ($p_j = p$);  \\ 
2) equal due dates $(d_j = d)$ and equal consumptions ($g_j = g$).

On the other hand, the special case of equal processing times ($p_j = p$) and ordered consumptions and due dates according to $g_1 \leq g_2 \leq \ldots \leq g_n$ and $d_1 \leq d_2 \leq \ldots \leq d_n$ is polynomially solvable. 

This paper considered also the budget scheduling problem with the objective to minimize the makespan. Instead of a resource consumption $g_j$ for each job, two values $g_j^-$ and $g_j^+$ are given, where $g_j^-$  has the same meaning as $g_j$ before. The value $g_j^+$ describes the additional earnings at the completion time of job $j$. The authors showed that the problem $1|NR, g_j^-, g_j^+|C_{max}$ belongs to the class $APX$ of problems allowing a constant-factor approximation algorithm. 

For future work, due to the $NP$-hardness of the majority of problems, it would be interesting to derive elimination  rules and structural properties of optimal solutions which can be used in enumerative algorithms.  

\section{ Concluding Remarks}  

In this paper, we reviewed some theoretical results on classical single machine scheduling problems. Since there exists a huge number of heuristic/metaheuristic and enumerative algorithms, 
our focus was on polynomial algorithms as well as complexity and approximability issues, discussed in the papers authored or co-authored by the authors of this work. The summarized polynomial special cases are also of interest when the conditions are not exactly satisfied and the resulting solution can be taken as a good heuristic one.

For single machine scheduling problems, efficient solution methods can often 
be found easier than for the corresponding multiprocessor or shop scheduling 
problems. However, such methods may give a better insight into the latter
more complex problems and may well serve as a basis for the development of
solution methods for these problems. At the same time, optimal solutions to
the single machine scheduling problems can directly be used to solve the  
corresponding multiprocessor or shop scheduling problems. For example, 
for such $NP$-hard problems, strong lower bounds can be obtained based on the
solutions of the corresponding single machine problems. Non-strict lower bounds,
which are easier and faster to obtain, can be used in any
search-tree based approximation algorithm, such as beam search, for example.
Similarly, using strict lower bounds, we obtain exact
branch-and-bound algorithms. Once implemented and tested, the
above algorithms can be used as an algorithmic engine for the
construction of decision support systems. Finally, we note that a number 
of algorithms for basic single machine problems can relatively easily be 
adopted using additional graph-completion mechanisms to solve the corresponding 
extended versions with precedence relations, transportation and setup times 
and other real-life job scheduling problems.

\end{document}